\documentclass[conference]{IEEEtran}

\usepackage{cite}
\usepackage{amsmath,amssymb,amsfonts}
\usepackage{graphicx}
\usepackage{booktabs}
\usepackage{subcaption}
\usepackage{textcomp}
\usepackage{xcolor}
\usepackage{multirow}
\usepackage{pifont}
\usepackage{algorithm}
\usepackage{algpseudocode}
\usepackage{url}
\usepackage[hidelinks]{hyperref}
\usepackage{tikz}
\usetikzlibrary{positioning, arrows.meta}

\newcommand{\sysname}{Zellige}

\newcounter{theorem}
\newcommand{\theoremhead}[1]{\noindent\refstepcounter{theorem}\textbf{Theorem~\thetheorem\ (#1).}}
\newcounter{proposition}

\title{\sysname{}: Moldable Sequence Placement for\\Mixed Image-Video DiT Training}

\author{
    \IEEEauthorblockN{
        Guangyu Xiang\textsuperscript{1,\textdagger},
        Xueze Kang\textsuperscript{1,\textdagger},
        Minwei Zhao\textsuperscript{1},
        Yuxin Wang\textsuperscript{2},
        Shaohuai Shi\textsuperscript{3},
        Lin Zhang\textsuperscript{4,*},
        Xiaowen Chu\textsuperscript{1,4,*}
    }
    \IEEEauthorblockA{
        \textsuperscript{1}HKUST (GZ);
        \quad \textsuperscript{2}No affiliation;
        \quad
        \textsuperscript{3}HIT (SZ);
        \quad \textsuperscript{4}HKUST
    }
    \IEEEauthorblockA{
        \{gxiang190,xkang507,mzhao886\}@connect.hkust-gz.edu.cn,
        yxwang.ele@gmail.com,\\
        shaohuais@hit.edu.cn,
        lzhangbv@connect.ust.hk,
        xwchu@ust.hk
    }
}

\begin{document}

\maketitle

\begingroup
\renewcommand{\thefootnote}{\fnsymbol{footnote}}
\footnotetext[2]{Equal contribution, \textsuperscript{*}Corresponding author.}
\endgroup

\begin{abstract}
High-quality video generation requires training Diffusion Transformers (DiTs) jointly on image and video data, posing a mixed-length sequence training problem across GPUs. Existing systems rely on data parallelism (DP), context parallelism (CP), or their combination; we model these designs as disjoint-group placement and prove that they face a fundamental tradeoff between inter-group load imbalance and intra-group communication redundancy. We present \sysname{}, a moldable sequence placement system that jointly selects each sequence’s parallelism configuration and participating ranks. \sysname{} consists of three components: a hardware profiler that estimates the execution time and memory consumption of candidate placements, a two-stage planner that balances compute-heavy anchor sequences and packs lighter filler sequences into the remaining capacity, and a coalesced attention engine that efficiently executes whole sequences alongside distributed-attention shards. Across 21 plans, the hardware profile predicts step makespan and peak allocated memory with mean absolute percentage errors of $3.4\%$ and $1.5\%$, respectively. The two-stage planner solves each batch in 33--119~ms, significantly faster than a joint-placement reference that optimizes all sequences together, while their modeled makespans differ by at most $0.32\%$. In end-to-end evaluations, \sysname{} outperforms KnapFormer by $1.12$--$1.48\times$ on 16 A800 GPUs and $1.27$--$1.54\times$ on 32 A6000 GPUs.

\end{abstract}


\section{Introduction}
\label{sec:intro-new}

Diffusion models have become a leading approach to high-quality video
generation~\cite{ddpm,latentdiffusion,videodiffusion}. Modern video generation
systems such as Wan~\cite{wan} and HunyuanVideo~\cite{hunyuanvideo} adopt
Diffusion Transformers (DiTs) as their denoising networks~\cite{dit}. During
training, each video is encoded into a spatiotemporal latent, which is then
corrupted with noise and divided into patches. The DiT processes these patches
as a token sequence and learns to predict the denoising target. The sequence
length grows with video duration and spatial resolution. For example, with
Wan's VAE compression and patch sizes, a 15-second 1080p video
produces a 497,760-token sequence~\cite{wan}.

Processing a minibatch containing such long sequences in a single training step
poses a systems challenge. In DiTs, dense self-attention makes per-sequence
compute grow quadratically with sequence length, while memory demand grows
approximately linearly~\cite{flashattention}. To meet these compute and memory requirements,
training systems distribute batch work across multiple GPUs at two basic
granularities. Data parallelism (DP) distributes work across samples by
assigning different complete sequences to GPUs and synchronizing gradients
after local computation~\cite{megatronflops}. Context parallelism (CP) instead
distributes work within a sample by partitioning one sequence's attention
computation across GPUs and communicating K/V states among them to compute exact
attention~\cite{loongtrain}.

Modern video DiT training commonly mixes images and videos across multiple
spatial and temporal resolutions to improve generation quality through
complementary appearance and motion supervision~\cite{wan,hunyuanvideo}.
These mixed batches span a wide range of sequence lengths, creating
inefficiencies for existing approaches built on either DP or CP.
Naive DP assigns complete sequences to balance total token counts across GPUs, but under
quadratic attention, one long video requires more compute than several images with the same total
tokens and can become a
straggler~\cite{hydraulis}. AdaptiveLoad~\cite{adaptiveload}
instead uses estimated sequence compute to adjust per-GPU batch sizes, but a long video remains
whole on one GPU and can still dominate that GPU's workload. To eliminate this
imbalance, USP~\cite{usp} applies the same all-rank CP configuration to every sequence, balancing attention
work across GPUs. However, this also makes short image sequences incur communication overhead even
though they could instead remain whole.

KnapFormer~\cite{knapformer} combines DP across rank groups with CP inside multi-rank groups to balance workloads while
limiting communication. It partitions the worker ranks into preconfigured
disjoint rank groups of possibly different sizes, then greedily assigns each sequence to one group
based on its estimated compute cost. This routing tends to place long sequences in larger
groups and short ones in smaller groups. However, because sequence work cannot cross group
boundaries, some groups may remain overloaded while others are underused. A short sequence in a
group using CP must also inherit that group's parallelism configuration and communicate even though
it could run whole on one rank. Our analysis proves that disjoint-group placement faces a fundamental
tradeoff between inter-group load imbalance and intra-group communication redundancy
(Sec.~\ref{sec:problem}).

These tradeoffs motivate more flexible sequence placement for mixed image-video DiT batches. The
planner treats each sequence as a moldable task~\cite{wuloiseau2023} whose parallelism configuration
and participating ranks are selected so that whole short sequences can share ranks with shards of
split long sequences. Supporting this co-location requires a low-latency per-batch planner and an
engine that executes whole sequences alongside distributed-attention shards.

To meet these requirements, we present \sysname{}, which combines a planner and execution engine to
provide moldable sequence placement for mixed image-video DiT training. 1) We formulate the placement
problem and analyze disjoint-group placement (Secs.~\ref{sec:new-formulation} and~\ref{sec:new-groups}),
proving that it faces a fundamental tradeoff between inter-group load imbalance and intra-group
communication redundancy (Secs.~\ref{sec:new-load-imbalance} and~\ref{sec:new-additional-splits}).
2) We develop a two-stage planner. Anchor Placement balances compute-heavy sequences (anchors) across ranks with
the OR-Tools CP-SAT solver~\cite{perron2023cpsat}, using an Exact Compact Formulation (ECF) to
aggregate solver-equivalent placements. Filler Packing keeps lighter sequences (fillers) whole and
places them in the remaining compute and memory headroom (Sec.~\ref{sec:new-anchor-solver}). 3) We
implement the Coalesced Attention Engine to efficiently execute placements in which whole sequences
share ranks with distributed-attention shards (Sec.~\ref{sec:new-implementation}).

We evaluate profile accuracy, planner efficiency, and end-to-end performance on two testbeds, one
with 16 NVIDIA Tesla A800 GPUs and the other with 32 NVIDIA RTX A6000 GPUs. Against GPU
measurements on 21 validation cases, the profile predicts step makespan and peak allocated memory
with mean absolute percentage error (MAPE) values of $3.4\%$ and $1.5\%$, respectively. In our
tests, planner solve time is 33--119~ms,
with at most $0.32\%$ step-makespan overhead relative to the joint-placement reference. In
end-to-end evaluation on mixed image-video workloads, \sysname{} is $1.12$--$1.48\times$ faster than
KnapFormer on 16 A800 GPUs and $1.27$--$1.54\times$ faster on 32 A6000 GPUs.

\section{Background and Motivation}
\label{sec:motivation}

\subsection{Video Diffusion Transformers (DiTs)}
\label{sec:dit-sequences}
\label{sec:quadratic}

\begin{figure}[t]
  \centering
  \begin{subfigure}[t]{0.51\columnwidth}
  \centering
  \includegraphics[width=\linewidth]{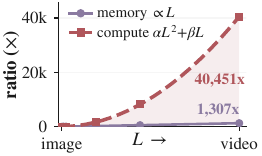}
  \caption{Modeled memory and compute demands.}
  \end{subfigure}\hfill
  \begin{subfigure}[t]{0.47\columnwidth}
  \centering
  \includegraphics[width=\linewidth]{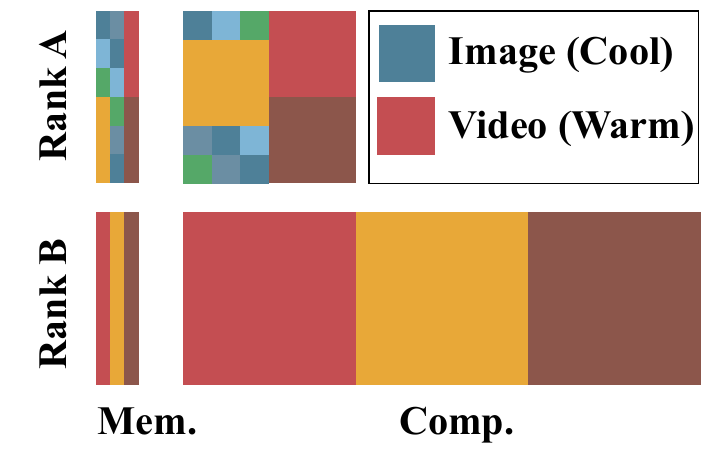}
  \caption{Equal tokens, unequal per-rank step load.}
  \end{subfigure}
  \caption{Token balance does not ensure compute balance. From a 256p image to a 10-second 1080p video,
  panel~(a) shows modeled memory and compute demands increasing by $1{,}307\times$ and
  $40{,}451\times$, respectively. Panel~(b) shows that equal token totals can
  carry unequal attention work, so the heavier rank determines the step makespan.}
  \label{fig:token-balance-gap}
  \end{figure}

During training, latent diffusion models use a denoising network to predict a denoising target from
the noisy latent representation of each image or video~\cite{latentdiffusion,wan,hunyuanvideo}.
Diffusion Transformers (DiTs) implement this network with Transformer blocks comprising dense
self-attention, projections, and feed-forward layers~\cite{dit}.
Each latent representation is partitioned into spatial or spatiotemporal patches, forming a token
sequence of length $L$. $L$ increases with spatial resolution and video duration. Projections and
feed-forward layers operate independently on each token and scale approximately linearly with $L$,
whereas dense self-attention evaluates $L^2$ token
pairs~\cite{megatronflops,kernelflume,adaptiveload}. Consequently, compute demand grows
quadratically with $L$, whereas memory demand under memory-efficient attention grows approximately
linearly~\cite{flashattention}.
Fig.~\ref{fig:token-balance-gap}(a) quantifies this divergence across the evaluated image and video
inputs, showing that modeled compute demand grows much faster than memory demand as $L$ increases.
Practical video inputs can produce sequences with hundreds of thousands of tokens, making multi-GPU
training necessary to handle the resulting compute and memory demands~\cite{wan}.

\begin{figure}[t]
  \centering
  \includegraphics[width=\linewidth]{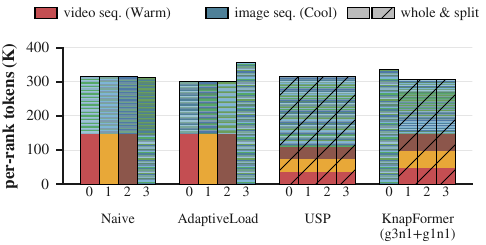}
  \caption{Per-rank token allocation for three 720p 10-second videos and 228 720p images on four
  ranks. Bars 0--3 stack each rank's sequences; warm and cool
  segments denote videos and images, while hatching denotes split shards. Naive and AdaptiveLoad
  keep sequences whole, USP splits every sequence across all ranks, and KnapFormer routes
  each sequence to one group in the \texttt{g3n1+g1n1} disjoint-group layout.}
  \label{fig:placement}
  \end{figure}
\subsection{Data and Context Parallelism}
\label{sec:dp-cp}

For sequence placement, multi-GPU DiT training distributes work at two granularities, across
complete sequences with data parallelism (DP) or within a sequence with context parallelism (CP).
Under DP, each rank processes
its assigned sequences before synchronizing gradients across
ranks~\cite{megatronflops}. Because complete sequences are not partitioned, DP cannot divide a long
sequence's attention work across ranks.

CP instead divides one sequence's attention among several ranks~\cite{loongtrain}. Each rank computes
a shard, while exact attention requires the ranks to exchange $K/V$ states from the full sequence.
This exchange introduces communication overhead, particularly when the participating ranks span
nodes. A query-sharded
all-gather design gathers remote $K/V$ blocks before each rank computes attention for its local
queries~\cite{llama3}.
Ulysses transforms sequence shards into head shards with all-to-all exchanges, performs local
attention, and exchanges the results back~\cite{ulysses}. Ring attention keeps $K/V$ blocks sharded
and circulates them around the ranks while accumulating the output~\cite{ringattention}.

\subsection{Existing Placement Approaches and Their Limitations}
\label{sec:existing-approaches}

Modern video DiTs jointly train on images and videos across spatial and temporal resolutions to
learn spatial appearance and temporal dynamics~\cite{wan,hunyuanvideo}. These mixed batches span a
wide range of sequence lengths. Because dense self-attention compute grows quadratically with sequence
length, ranks with equal token totals can still carry unequal compute loads, as
Fig.~\ref{fig:token-balance-gap}(b) illustrates. Rank B receives longer video sequences and therefore
carries more attention work than Rank A despite the same token total.

\begin{figure}[t]
  \centering
  \begin{subfigure}[t]{0.5\columnwidth}
    \centering
    \includegraphics[width=\linewidth]{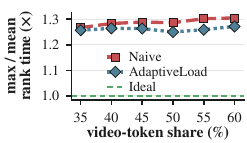}
    \caption{Naive and AdaptiveLoad imbalance.}
    \label{fig:keepwhole-rank-imbalance}
  \end{subfigure}%
  \begin{subfigure}[t]{0.5\columnwidth}
    \centering
    \includegraphics[width=\linewidth]{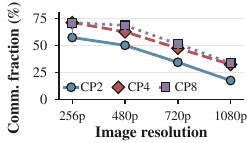}
    \caption{Communication fraction.}
    \label{fig:communication-fraction-panel}
  \end{subfigure}
  \caption{Rank imbalance and CP communication cost.
  Panel~(a) reports rank-time imbalance (max/mean) for Naive and AdaptiveLoad on four
  A800 GPUs across six video-token shares;
  $1.0\times$ denotes ideal balance. Panel~(b) reports the share of attention time spent on
  communication when processing images within a training step under multiple CP degrees.}
  \label{fig:communication-fraction}
\end{figure}

Existing DiT training approaches build on DP, CP, or their combination.
Fig.~\ref{fig:placement} shows how existing approaches place three 720p 10-second videos and 228
720p images across four ranks. Naive DP assigns each image or video sequence to one rank to equalize
total token counts across ranks. In Fig.~\ref{fig:placement}, it places one video on each of
ranks 0--2 and only images on rank 3, producing similar token totals but heavier compute loads on
ranks 0--2.
AdaptiveLoad makes DP compute-aware by dynamically adjusting each rank's batch size
according to estimated sequence compute~\cite{adaptiveload}. Fig.~\ref{fig:placement} shows the
resulting placement under the per-rank memory cap, with more images on rank 3 and fewer on each video
rank. However, each long video must still run on a single rank and can remain a straggler.
Fig.~\ref{fig:communication-fraction}(a) extends this case to six video-token shares from 35\% to
60\% and measures full-step rank-time imbalance on four A800 GPUs.
Across these workloads, Naive produces a $1.27$--$1.31\times$ max/mean rank-time ratio, whereas
AdaptiveLoad lowers the ratio only to $1.25$--$1.27\times$, leaving the slowest rank
$25$--$27\%$ above the mean. Both therefore remain far from the ideal $1.0\times$ balance.

USP arranges the CP ranks in a Ulysses-by-ring grid, using all-to-all exchanges along the Ulysses
dimension and ring communication along the ring dimension~\cite{usp}. It applies the same static CP
configuration to every sequence in the batch. In Fig.~\ref{fig:placement}, this configuration
splits every video and image across all four ranks. This balances long-video work but imposes
communication and buffer-memory overhead on short images that could instead remain whole.
Fig.~\ref{fig:communication-fraction}(b) measures the share of attention time spent on
communication when processing images within a training step. Communication accounts for
$57.3$--$71.0\%$ of attention time at 256p and $17.7$--$33.8\%$ at 1080p, with larger CP groups
generally increasing the fraction.

KnapFormer combines DP across rank groups with CP inside multi-rank groups~\cite{knapformer}. It partitions
ranks into preconfigured disjoint Ulysses-style rank groups of possibly different sizes and
greedily routes each sequence by a $\gamma$-corrected compute cost into exactly one group.
In Fig.~\ref{fig:placement}, the \texttt{g3n1+g1n1} disjoint-group layout uses rank 0 as a singleton group and
ranks 1--3 as a three-rank group. This permits whole and three-rank execution within one step, but
the group boundaries prevent work from being shared across groups and can leave one overloaded
while the other is underused. Moreover, images routed to the three-rank group inherit its split and
must communicate even though they could run whole on one rank.
Secs.~\ref{sec:new-load-imbalance} and~\ref{sec:new-additional-splits} formalize this fundamental
tradeoff between inter-group load imbalance and intra-group communication redundancy.

\noindent\textbf{The gap.}
Mixed image-video DiT batches therefore need more flexible per-sequence placement. Selecting a
parallelism configuration and participating ranks for each sequence allows rank sets to overlap, so
whole short sequences can share ranks with long-video shards. These choices must be coordinated
under the step-makespan objective and per-rank memory cap.

\section{Problem Formulation and the Limits of Disjoint-Group Placement}
\label{sec:problem}

\subsection{Placement Problem}
\label{sec:new-formulation}

For one synchronized step, let $R=\{1,\ldots,N\}$ be the worker ranks and $S$ the minibatch
sequences. Each $s$ has a set $\mathcal C_s$ of legal executable placement options. An option
$c=(\theta_c,R_c)\in\mathcal C_s$ pairs a parallelism configuration
$\theta_c=(q_c,h_c,k_c)$ over the query, head, and key/value axes with a participating-rank set
$R_c$. The configuration fixes the primitive composition and factorization before rank assignment,
and its split degree is $\lambda_c=q_ch_ck_c=|R_c|$. A placement $x$ selects one option for every sequence, and the action
space $\mathcal X$ is the set of placements admitted by the policy under consideration.
Using the same per-sequence option menus, the matched joint relaxation is
$\mathcal X_{\mathrm{joint}}:=\prod_{s\in S}\mathcal C_s$, so each sequence selects any legal option
independently without being restricted to shared disjoint groups.

For placement $x$, binary $x_{s,c}$ indicates whether sequence $s$ selects option $c$. The
option contributes time $A_{s,c,r}$ and allocated memory $M_{s,c,r}$ to rank $r$, both zero
outside $R_c$. The time on rank $r$ is the sum of the time contributions from its selected options. Since
one synchronized step completes only after all ranks finish, its step makespan is the
largest rank time. The resulting optimization is
\begin{equation}
  \begin{aligned}
  T_{\mathcal X}^{\star}
  &=\min_{x\in\mathcal X}\max_{r\in R}
    \sum_{\substack{s\in S\\c\in\mathcal C_s}}x_{s,c}A_{s,c,r},\\
  \mathrm{s.t.}\qquad
  &\sum_{c\in\mathcal C_s}x_{s,c}=1
    &&\forall s,\\
  &x_{s,c}\in\{0,1\}
    &&\forall s,\ c\in\mathcal C_s,\\
  &\sum_{\substack{s\in S\\c\in\mathcal C_s}}x_{s,c}M_{s,c,r}\le M_{\mathrm{cap}}
    &&\forall r.
  \end{aligned}
  \label{eq:new-makespan}
\end{equation}
The objective selects a placement in $\mathcal X$ that minimizes the step makespan.
The first two constraint lines enforce one binary option choice for every sequence, and the third
keeps each rank's allocated memory within $M_{\mathrm{cap}}$.

\begin{table}[t]
  \centering
  \caption{Notation for placement and theoretical bounds.}
  \label{tab:new-notation}
  \footnotesize
  \setlength{\tabcolsep}{3.5pt}
  \renewcommand{\arraystretch}{1.04}
  \begin{tabular}{@{}p{0.315\columnwidth}p{0.605\columnwidth}@{}}
    \toprule
    Symbol & Meaning \\
    \midrule
    $R,N;\ S,s$ & worker ranks and minibatch sequences \\
    $\mathcal C_s,c$ & set of legal executable placement options and one executable placement option \\
    $\theta_c$ & parallelism configuration $(q_c,h_c,k_c)$ \\
    $\lambda_c,R_c$ & split degree and participating-rank set \\
    $x,x_{s,c}$ & placement and binary option-selection variable \\
    $A_{s,c,r},M_{s,c,r}$ & per-rank time and memory contributions \\
    $M_{\mathrm{cap}}$ & memory cap \\
    $\mathcal X,T_{\mathcal X}^{\star}$ & action space and minimum step makespan \\
    $\mathcal X_{\mathrm{DP}},\mathcal X_{\mathrm{CP}}$ & pure DP and pure CP action spaces \\
    $\mathcal X_{\mathrm{disj}}(\mathcal G),\mathcal X_{\mathrm{disj}},\mathcal X_{\mathrm{joint}}$ & disjoint-group under $\mathcal G$, complete disjoint-group, and joint action spaces \\
    $\mathcal G;(G_j,\theta_j)$ & disjoint rank groups with fixed configurations; one group/configuration pair \\
    $g_j,J,g_{\max}$ & group size, group count, and largest group size \\
    $\mathcal Y(\mathcal G),y_{s,j}$ & feasible assignments and binary variable \\
    $\mathbf w,w_s,W,w_{\max},\phi$ & work vector, sequence/total/maximum work, and concentration \\
    $T_{\mathrm{disj}}^{\star}(\mathcal G;\mathbf w),T_{\mathrm{joint}}^{\star}$ & disjoint-group and joint minimum compute makespans \\
    $\Gamma(\mathbf w,\mathcal G)$ & ratio of the two compute makespans \\
    \bottomrule
  \end{tabular}
\end{table}

\subsection{Disjoint-Group Placement}
\label{sec:new-groups}

Disjoint-group placement restricts $\mathcal X$ through
$\mathcal G=\{(G_1,\theta_1),\ldots,(G_J,\theta_J)\}$, where the rank sets are disjoint and cover $R$:
$\bigcup_jG_j=R$ and $G_i\cap G_j=\varnothing$ for $i\ne j$. Each group has size
$g_j=|G_j|$ and a fixed legal parallelism configuration
$\theta_j=(q_j,h_j,k_j)$ satisfying $q_jh_jk_j=g_j$. For a fixed $\mathcal G$,
$\mathcal X_{\mathrm{disj}}(\mathcal G)$ contains placements in which each sequence enters exactly
one group and selects the option $c=(\theta_j,G_j)\in\mathcal C_s$ associated with that group. A singleton group
keeps its sequences whole, whereas a multi-rank group applies its configuration across all $g_j$
ranks, with $\lambda_c=g_j$ and $R_c=G_j$.
KnapFormer's preconfigured disjoint rank groups instantiate this restriction; each multi-rank group uses a Ulysses
configuration whose degree equals the group size~\cite{knapformer}. Across all legal $\mathcal G$, the
complete disjoint-group action space is
$\mathcal X_{\mathrm{disj}}=\bigcup_{\mathcal G}\mathcal X_{\mathrm{disj}}(\mathcal G)$. This
union includes every legal choice of multiple, possibly unequal disjoint groups.

The disjoint-group family contains pure DP and pure CP as endpoints. Pure DP keeps every sequence
on one rank, with
$\mathcal X_{\mathrm{DP}}:=\mathcal X_{\mathrm{disj}}(\{(\{j\},(1,1,1))\}_{j=1}^{N})$.
Pure CP splits every sequence across all $N$ ranks under one legal configuration, with
$\mathcal X_{\mathrm{CP}}:=\bigcup_{\theta}\mathcal X_{\mathrm{disj}}(\{(R,\theta)\})$, where the
union ranges over legal $\theta=(q,h,k)$ satisfying $qhk=N$. By construction, both are subsets of
$\mathcal X_{\mathrm{disj}}$.

\subsection{Inter-Group Load Imbalance}
\label{sec:new-load-imbalance}

For a given $\mathcal G$, we compare optimal placements in
$\mathcal X_{\mathrm{disj}}(\mathcal G)$ and the matched joint relaxation
$\mathcal X_{\mathrm{joint}}$. The joint relaxation can reproduce every disjoint-group assignment
under $\mathcal G$ and may also use overlapping rank sets. Hence
$\mathcal X_{\mathrm{disj}}(\mathcal G)\subseteq\mathcal X_{\mathrm{joint}}$, with strict containment
possible. Comparing their optima isolates the disjoint-group restriction.

We use an ideal compute-only model with $N\ge2$, zero communication cost, and nonbinding memory.
Every option conserves its sequence's total work, and every sequence can be evenly split across the
ranks of any group or across all $N$ ranks. Let $\mathbf w=(w_s)_{s\in S}$ be the work vector, with $w_s>0$. For
this $\mathcal G$, binary $y_{s,j}\in\{0,1\}$ assigns each sequence to exactly one group, and
$\mathcal Y(\mathcal G)$ denotes the assignments satisfying $\sum_jy_{s,j}=1$. Assigning $s$ to
$G_j$ spreads $w_s$ evenly across its $g_j$ ranks, so the disjoint-group optimum under $\mathcal G$ is
\begin{equation}
  T_{\mathrm{disj}}^{\star}(\mathcal G;\mathbf w)
  =\min_{y\in\mathcal Y(\mathcal G)}\max_j\frac{\sum_sy_{s,j}w_s}{g_j}.
  \label{eq:new-batch-disj-opt}
\end{equation}
Let $W=\sum_sw_s$, $w_{\max}=\max_sw_s$, $g_{\max}=\max_jg_j$, and
$\phi=w_{\max}/W$. Assigning every sequence to an all-rank option attains the average-load lower
bound $W/N$, so $T_{\mathrm{joint}}^{\star}=W/N$. We compare the two optima using the ratio
$\Gamma(\mathbf w,\mathcal G)
=T_{\mathrm{disj}}^{\star}(\mathcal G;\mathbf w)/T_{\mathrm{joint}}^{\star}
=NT_{\mathrm{disj}}^{\star}(\mathcal G;\mathbf w)/W$.

\theoremhead{Inter-group load imbalance}\label{thm:group-load-imbalance}
For every $\mathcal G$ and every finite positive-work batch,
\begin{equation}
  \max\!\left\{1,\ \frac{N}{g_{\max}}\phi\right\}
  \le \Gamma(\mathbf w,\mathcal G)
  \le \frac{N}{g_{\max}}.
  \label{eq:new-load-imbalance-bounds}
\end{equation}
For each $\mathcal G$, the exact worst-case ratio over all finite positive-work batches under that
$\mathcal G$ is $N/g_{\max}$.
\emph{Proof.}
The average load gives $\Gamma\ge1$, while the largest sequence occupies at most $g_{\max}$ ranks and
gives $\Gamma\ge N\phi/g_{\max}$. Assigning the complete batch to a largest group gives
$\Gamma\le N/g_{\max}$. A one-sequence workload makes the lower and upper bounds equal to
$N/g_{\max}$, proving the worst-case equality.
\hfill$\square$

Equality $\Gamma(\mathbf w,\mathcal G)=1$ holds iff some $y\in\mathcal Y(\mathcal G)$ assigns each
group $G_j$ exactly $g_jW/N$ work. Hence whether a batch attains perfect load balance under
$\mathcal G$ depends on this proportional packing, whereas $N/g_{\max}$ characterizes only the
worst case across batches. Guaranteeing $\Gamma=1$ for every batch under a given $\mathcal G$
requires $g_{\max}=N$, which
yields the single all-rank group of pure CP and splits every sequence.

\subsection{Intra-Group Communication Redundancy}
\label{sec:new-additional-splits}

Under the same ideal compute-only model, a sequence is split and invokes communication when
$\lambda_c>1$. Among placements with the same makespan, intra-group communication redundancy is the
additional split count under disjoint-group placement relative to joint placement. Each filler has
a legal whole-sequence option on every rank, and all sequences share one legal all-rank
configuration that divides their work evenly. For
any integer $p\ge1$, take one giant of work $Nu$ and $m=Np$ fillers of work $\delta>0$. Choose
$u>(N-1)p\delta$ and set $F=u+p\delta<Nu/(N-1)$.

\theoremhead{Intra-group communication redundancy}\label{thm:additional-splits}
Both action spaces attain the average-load lower bound $F$. Among placements with makespan $F$, the
minimum split count is $m+1$ under $\mathcal X_{\mathrm{disj}}$ but one under
$\mathcal X_{\mathrm{joint}}$. Thus, disjoint-group placement incurs intra-group communication
redundancy of $m=Np$, which is unbounded for fixed $N$.

\emph{Proof.}
The total work is $Nu+m\delta=NF$, so the makespan is at least $F$. Joint placement attains $F$ by
splitting the giant $N$ ways and placing $p$ whole fillers per rank; $Nu>F$ makes that split
necessary. For a disjoint-group placement attaining $F$, the giant's group size $g$ cannot be less
than $N$, since
$Nu/g\ge Nu/(N-1)>F$. Thus $g=N$; because the groups are disjoint, the all-rank group leaves no
ranks for another nonempty group, so all $m$ fillers join it and split. This placement gives per-rank work
$u+m\delta/N=F$, so $m+1$ is minimal. Letting $p$ grow with $N$ fixed proves that intra-group
communication redundancy is unbounded.
\hfill$\square$

\section{\sysname{} Design}
\label{sec:solve}

\begin{figure}[t]
  \centering
  \includegraphics[width=\columnwidth]{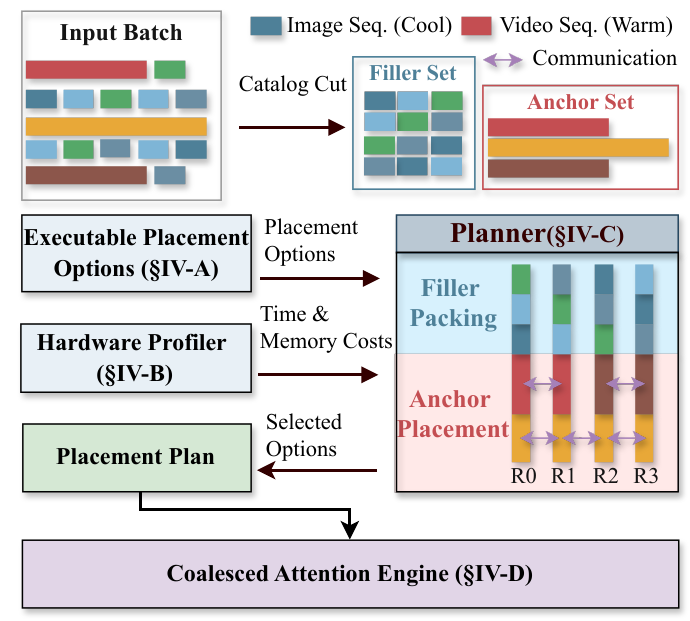}
  \caption{\sysname{} workflow. A catalog cut partitions the batch into anchor and filler
  sets. Given executable placement options, profiled execution times for their parallelism
  configurations, and analytical allocated-memory costs, Anchor Placement selects one option per anchor, and
  Filler Packing assigns whole fillers to the remaining rank
  headroom. The Coalesced Attention Engine executes the resulting placement; lavender arrows mark
  the planned collective communication of split anchors.}
  \label{fig:zellige-overview}
\end{figure}
\sysname{} realizes moldable sequence placement with a two-stage planner and the Coalesced Attention Engine.
It enumerates each sequence's executable placement options and runs the Hardware Profiler once, then plans and executes each batch under the
per-rank memory cap, allowing split anchors and whole fillers to share ranks
(Fig.~\ref{fig:zellige-overview}).

\subsection{Executable Placement Options}
\label{sec:new-menu}

Disjoint-group placement locks each sequence into one rank group, so the shards of a split giant can
never share ranks with whole short sequences~\cite{knapformer,flexsp}. In contrast, \sysname{}
constructs a set of executable placement options for each sequence by pairing a parallelism
configuration $(q,h,k)$ with a legal participating-rank set; rank sets may overlap across sequences.
Here, $q$, $h$, and $k$ are the query-, head-, and key/value-axis split factors, implemented by
all-gather~\cite{llama3}, Ulysses~\cite{ulysses}, and ring attention~\cite{ringattention},
respectively, with split degree $\lambda=qhk$. The configuration $(1,1,1)$ keeps the sequence whole
and activates no attention communication. Table~\ref{tab:new-menu} summarizes the communication and
per-rank resident $K/V$ payloads of these configurations. Multi-axis configurations combine these
primitives, but only profiled configurations satisfying executor-specific legality constraints are
retained; for example, a Ulysses degree must divide the number of attention heads~\cite{ulysses}.

\subsection{Hardware Profiler}
\label{sec:new-cost}

The Hardware Profiler supplies the time and memory costs used by the planner. It runs once per
training setup and, following profile-guided scheduling practice~\cite{perfmodel,fsmoe}, measures the
per-rank execution time of every supported bucket--configuration pair. Each measurement captures
attention compute, communication, kernel fusion, launch overhead, overlap, and
packing~\cite{compass,mgwfbp,dear}, retaining primitive composition, degree, factorization, and
executor effects without assuming separability. After removing a placement-independent step cost,
the profiler stores the resulting per-participating-rank price $P_{b,\theta}$ in $\mathcal P$. With
$b(s)$ denoting sequence $s$'s catalog bucket, the planner instantiates Eq.~\eqref{eq:new-makespan} by setting
$A_{s,c,r}=P_{b(s),\theta_c}$ for $r\in R_c$ and zero otherwise, so each rank load is the sum of its
selected prices. The removed cost is added back when predicting the step makespan.
We evaluate the additive model in Sec. V-B.
\begin{table}[t]
  \centering
  \caption{The three base primitives, one per partition axis. A composed parallelism configuration
  combines the communication required by its active axes, while the whole-sequence configuration
  $(1,1,1)$ activates none. Ulysses' two exchanges sit
  before and after attention. The communication entries show the collective pattern and payload size,
  while the last column reports the resident $K/V$ payload per rank for a length-$L$ sequence.
  Transient collective workspaces are accounted for separately in $M^{\mathrm{buf}}$, where $kv$
  denotes per-token key-value bytes.}
  \label{tab:new-menu}
  \footnotesize
  \setlength{\tabcolsep}{3pt}
  \begin{tabular}{@{}llll@{}}
    \toprule
    Primitive & Configuration & Communication & Resident $K/V$ \\
    \midrule
    all-gather & $(q,1,1)$ & gather $K/V$ once, $L\,kv$ & replicated, $L\,kv$ \\
    Ulysses    & $(1,h,1)$ & head all-to-all $\times 2$, $L\,kv/h$ & sharded, $L\,kv/h$ \\
    ring       & $(1,1,k)$ & stream $K/V$ $\times(k{-}1)$, $L\,kv/k$ & sharded, $L\,kv/k$ \\
    \bottomrule
  \end{tabular}
\end{table}

For memory, the profiler uses an analytical model based on resident-tensor shapes and
lifetimes~\cite{xema}. On a participating rank,
$M_{s,c,r}=M^{\mathrm{act}}_{s,c,r}+M^{\mathrm{KV}}_{s,c,r}+M^{\mathrm{buf}}_{s,c,r}$, and the cost is
zero otherwise. $M^{\mathrm{act}}$, $M^{\mathrm{KV}}$, and $M^{\mathrm{buf}}$ denote resident
activations, resident $K/V$ payloads, and collective communication buffers, respectively.
$M^{\mathrm{act}}$ depends on model geometry, depth, checkpointing mode, and local shard
length~\cite{megatronsp}. The communication terms depend on how the primitive stores and
moves $Q/K/V$. All-gather replicates the full resident $K/V$ payload, of size $L\,kv$, on every
participating rank, with transient gather workspace counted in $M^{\mathrm{buf}}$. Ulysses reduces
the resident $K/V$ payload to $L\,kv/h$ through head sharding,
while its QKV-packed pre-attention all-to-all and post-attention inverse all-to-all create transient
input/output buffers. Ring attention retains an $L\,kv/k$ local shard and streams remote blocks
through transient send/receive buffers. For multi-axis configurations, the model follows the
executor's nested tensor shapes and lifetimes rather than summing the single-axis entries.
Calibration against measured peak allocated memory accounts for implementation-specific workspaces
and supplies $M_{\mathrm{base}}$ and $M_{\mathrm{usable}}$, giving
$M_{\mathrm{cap}}=M_{\mathrm{usable}}-M_{\mathrm{base}}$ for Eq.~\eqref{eq:new-makespan}.

\subsection{Anchor/Filler Planner}
\label{sec:new-anchor-solver}

\sysname{} applies a fixed catalog-level cut that assigns each batch sequence to the anchor set
or the filler set, then plans their placement in two stages. Anchor sequences are
typically long videos with high execution costs and a low communication fraction, so Anchor
Placement can split them across ranks to mitigate stragglers. Filler sequences are typically
lightweight images or short clips, so splitting them adds high communication overhead relative to
their compute. Filler Packing therefore keeps them whole and places them within
the compute and memory headroom left by the anchors.

\sysname{} orders the catalog buckets as $b_{(1)},\ldots,b_{(B_{\mathrm{cat}})}$ by nondecreasing
profiled whole-sequence time $w_b=P_{b,\mathrm{whole}}$, and writes $w_i=w_{b_{(i)}}$. It selects
the contiguous cut that minimizes the within-set dispersion of $\log w_i$~\cite{fisher1958},
\begin{equation}
  \begin{split}
  k^\star=\arg\min_{1\le k<B_{\mathrm{cat}}}\biggl[
  &\sum_{i\le k}(\log w_i-\mu_{\mathrm{fill}}(k))^2 \\
  &+\sum_{i>k}(\log w_i-\mu_{\mathrm{anc}}(k))^2\biggr],
  \end{split}
  \label{eq:new-catalog-cut}
\end{equation}
where $\mu_{\mathrm{fill}}(k)$ and $\mu_{\mathrm{anc}}(k)$ are the mean log scores on either side;
ties choose the smallest $k$. The logarithm captures multiplicative separation and is invariant to
common price rescaling. Sequence $s$ is an anchor iff $b(s)=b_{(i)}$ for some $i>k^\star$;
otherwise, it is a filler.

\noindent\textbf{Stage 1: Anchor Placement.}
Stage~1 chooses one executable placement option for each anchor, thereby fixing its primitive
composition, split degree, and participating ranks.
Anchor Placement uses an $\epsilon$-constrained secondary objective~\cite{haimes1971epsilon}. It first minimizes the maximum profiled anchor load under the
per-rank memory cap, then minimizes the number of split anchors while allowing that optimum to increase
by at most $\epsilon$. This keeps the balance bound explicit and avoids a conversion weight between
maximum profiled anchor load and split count. Choices of parallelism configuration and rank set are discrete,
whereas profiled rank load, analytical memory, and split count are linear in those choices. This
discrete-linear structure fits CP-SAT's Boolean and bounded-integer variables and linear constraints.
We therefore use OR-Tools CP-SAT~\cite{perron2023cpsat}, which reports when the integer optimum is
certified.

A direct CP-SAT model uses one binary variable per modeled sequence--option pair, as in
Eq.~\eqref{eq:new-makespan}. Under DiT's small, fixed bucket catalog, many such variables share
placement options and resource coefficients and differ only by sequence identity, creating
permutation-equivalent solutions~\cite{margot2003}. Building on CP-SAT, we use an \emph{Exact Compact Formulation
(ECF)} to replace these interchangeable binaries with bounded integer counts; we call the resulting
implementation CP-SAT-ECF.

We group solver-equivalent sequences in the set being optimized and denote the resulting set of
groups by $\mathcal E$. Members of the same group have the same bucket and length, executable
placement options, and coefficients used to compute load and memory.
Group $g\in\mathcal E$ contains $n_g$ sequences. For each group $g$, let $j\in\mathcal J_g$ index a
parallelism configuration, with degree $\lambda_{g,j}$, profiled
time price $p_{g,j}$ and analytical memory $m_{g,j}$ per
participating rank, and split indicator
$\xi_{g,j}=\mathbf 1\{\lambda_{g,j}>1\}$. Let $\mathcal B_{g,j}$ denote the legal participating-rank
sets for this group and configuration. Each $B\in\mathcal B_{g,j}$ contains exactly
$\lambda_{g,j}$ distinct ranks, so $(j,B)$ identifies an executable placement option for members of $g$.
ECF defines $z_{g,j}$ as the number of group-$g$ sequences choosing parallelism configuration $j$ and $u_{g,j,B}$ as the
number assigned to rank set $B$. Its bounded-integer constraints are
\begin{equation}
  \begin{gathered}
  z_{g,j},u_{g,j,B}\in\{0,\ldots,n_g\},\\[-0.2em]
  \sum_{j\in\mathcal J_g}z_{g,j}=n_g\quad\forall g,
  \qquad
  \sum_{B\in\mathcal B_{g,j}}u_{g,j,B}=z_{g,j}\quad\forall g,j.
  \end{gathered}
  \label{eq:new-ecf-feasible}
\end{equation}
From $u$, we derive $\eta_{g,j,r}(u)$, the number of group-$g$ sequences choosing parallelism configuration $j$ that use rank
$r$. The resulting rank load, memory, and split count are
\begin{equation}
  \begin{alignedat}{2}
  \eta_{g,j,r}(u)&=\sum_{\substack{B\in\mathcal B_{g,j}\\r\in B}}u_{g,j,B},
  &\quad L_r&=\sum_{g,j}p_{g,j}\eta_{g,j,r}(u),\\[-0.15em]
  H_r&=\sum_{g,j}m_{g,j}\eta_{g,j,r}(u),
  &\quad K&=\sum_{g,j}\xi_{g,j}z_{g,j}.
  \end{alignedat}
  \label{eq:new-ecf-resources}
\end{equation}
Let $\mathcal F_{\mathrm{ECF}}$ denote the integer assignments satisfying
Eq.~\eqref{eq:new-ecf-feasible}. To express the maximum-rank objective of
Eq.~\eqref{eq:new-makespan} with linear constraints, we introduce $T$ and require $L_r\le T$ for
every rank. Anchor Placement performs the following balance and split-count solves.
\begin{equation}
  \begin{aligned}
  T_{\mathrm{anc}}^\star
  &=\min_{(z,u)\in\mathcal F_{\mathrm{ECF}},\,T\ge0} T\\
  &\text{s.t. } L_r\le T,\quad H_r\le M_{\mathrm{cap}}\qquad\forall r .
  \\[-0.2em]
  K_{\mathrm{anc}}^\star
  &=\min_{(z,u)\in\mathcal F_{\mathrm{ECF}}} K\\
  &\text{s.t. } L_r\le(1+\epsilon)T_{\mathrm{anc}}^\star,\quad
    H_r\le M_{\mathrm{cap}}\qquad\forall r .
  \end{aligned}
  \label{eq:new-ecf-objectives}
\end{equation}
We set $\epsilon=10^{-3}$, so the second solve permits at most $0.1\%$ excess over a certified
primary optimum. An empty anchor set has the trivial optima $T=K=0$ and proceeds to Filler Packing.
When both rounds return \textsc{Optimal}, they certify Anchor Placement's in-model balance optimum
and split-count optimum within the $\epsilon$ load slack, under the specified options,
integer-scaled costs, and memory cap.

\noindent\textbf{ECF exactness.}
When ECF and the per-sequence formulation use the same executable placement options and additive
per-rank time and memory model, ECF preserves
the feasible rank-load and memory vectors and split count of the per-sequence formulation.
Aggregating any concrete placement by group, configuration,
and rank set yields a feasible $(z,u)$. Conversely, the constraints in
Eq.~\eqref{eq:new-ecf-feasible} partition the $n_g$ solver-equivalent members of each group
among legal options $(j,B)$. Because members within a group have identical coefficients, expanding
these counts yields a legal placement with the same
$(L_1,\ldots,L_N,H_1,\ldots,H_N,K)$; only permutations of members within a group are removed.
The ECF variable count depends on group signatures and placement-option structure, not group
multiplicities $n_g$.

\noindent\textbf{Stage 2: Filler Packing.}
Stage~2 initializes its current per-rank load and memory ledgers from the decoded anchor placement,
$C_r\gets L_r$ and $M_r\gets H_r$. It keeps every sequence in $\mathcal S_{\mathrm{fill}}$ whole
and processes them in decreasing profiled whole-sequence price
$p_s=P_{b(s),\mathrm{whole}}$. Define the current compute and memory headroom as
$h^C_r=F-C_r$ and $h^M_r=M_{\mathrm{cap}}-M_r$, where
$F=\sum_{s\in S}P_{b(s),\mathrm{whole}}/N$ is the profiled zero-communication load floor. For filler
sequence $s$, Filler Packing chooses
\begin{equation}
  \begin{aligned}
  r^\star(s)
  &=\arg\max_{r\in\mathcal R_s}\min\{\rho^C_{s,r},\rho^M_{s,r}\},\\
  \rho^C_{s,r}
  &=\frac{h^C_r-p_s}{F},\qquad
  \rho^M_{s,r}=\frac{h^M_r-M_{s,\mathrm{whole},r}}{M_{\mathrm{cap}}},\\
  \mathcal R_s
  &=\{r:h^M_r\ge M_{s,\mathrm{whole},r}\}.
  \end{aligned}
  \label{eq:new-balanced-fill}
\end{equation}
For each feasible rank, the score uses the smaller post-placement compute and memory slack as the
bottleneck. Maximizing this bottleneck places $s$ where the tighter resource retains the most headroom,
so neither compute nor memory is balanced at the expense of the other. After assigning each filler, the planner
updates $C_{r^\star}$ and $M_{r^\star}$, then recomputes the feasible ranks and scores for the next
filler sequence. \sysname{} returns
a placement only when both stages succeed; otherwise, planning fails.
Algorithm~\ref{alg:zellige} summarizes this two-stage procedure.

\begin{algorithm}[t]
  \caption{The \sysname{} two-stage planner}
  \label{alg:zellige}
  \begin{algorithmic}[1]
    \Require $S$, $k^\star$, $\{\mathcal C_s\}$, $R$, $\mathcal P$, $\{M_{s,c,r}\}$,
      $M_{\mathrm{cap}}$, $\epsilon$
    \State $(\mathcal S_{\mathrm{anc}},\mathcal S_{\mathrm{fill}})\gets\operatorname{ApplyCatalogCut}(S,k^\star)$
    \State $\mathcal E\gets\operatorname{GroupEquivalentAnchors}(\mathcal S_{\mathrm{anc}})$
    \State $(T_{\mathrm{anc}}^\star,\sigma_1)\gets
      \operatorname{ECFBalance}(\mathcal E,M_{\mathrm{cap}})$
    \State \textbf{if} $\sigma_1\ne\textsc{Optimal}$ \textbf{then return failed}
    \State $\overline T_{\mathrm{anc}}\gets(1+\epsilon)T_{\mathrm{anc}}^\star$
    \State $(z,u,\sigma_2)\gets
      \operatorname{ECFMinSplit}(\mathcal E,\overline T_{\mathrm{anc}},M_{\mathrm{cap}})$
    \State \textbf{if} $\sigma_2\ne\textsc{Optimal}$ \textbf{then return failed}
    \State $x\gets\operatorname{DecodeECF}(\mathcal E,z,u)$
    \State initialize $C_r\gets L_r$, $M_r\gets H_r$; set $F\gets\sum_{s\in S}P_{b(s),\mathrm{whole}}/N$
    \For{$s\in\mathcal S_{\mathrm{fill}}$ in decreasing $P_{b(s),\mathrm{whole}}$}
      \State compute $\mathcal R_s$; \textbf{if} empty \textbf{then return failed}; otherwise compute $r^\star(s)$
      \State assign $s$ to the whole-sequence option on $r^\star(s)$ in $x$
      \State update current rank load and memory
    \EndFor
    \State \Return $x$, $(\sigma_1,\sigma_2)$, \textsc{Success}
  \end{algorithmic}
\end{algorithm}

\subsection{Coalesced Attention Engine}
\label{sec:new-implementation}

A placement plan may assign a rank both whole sequences and shards from distributed-attention
groups. Executing segments separately would incur many small kernel launches and collectives. The
Coalesced Attention Engine lowers the placement into a rank-local schedule and coalesces compatible
segments without mixing attention across sequences.

Given the placement, each rank uses the sequences' global token ranges to materialize only its assigned
tokens and bind them to the selected attention executors. Process groups are cached by rank set for reuse
across steps. The engine then stably orders segments by decreasing split degree and uses
the global placement order to break ties. Placing wider collectives first avoids serializing work on
disjoint rank groups, while ensuring that participating ranks invoke collectives in the same order.
Within this schedule, the engine coalesces compatible work in two ways. First, whole sequences assigned
to the same rank are packed into one block-diagonal FlashAttention call. Second, pure Ulysses sequences
sharing the same configuration and rank set are batched into one QKV-packed all-to-all, one
block-diagonal attention call, and one inverse all-to-all. Other segments use their selected executors
independently, while ring-based executors retain their native communication--attention overlap.
Block-diagonal attention keeps sequences separate~\cite{navit} and returns outputs in the same rank-local order as
the inputs.

\section{Evaluation}
\label{sec:eval}

\subsection{Experimental Setup}
\label{sec:new-eval-setup}

\noindent\textbf{Testbeds, model, and workloads.}
We use two multi-node GPU testbeds: 16 NVIDIA Tesla A800 GPUs across two eight-GPU nodes and 32
NVIDIA RTX A6000 GPUs across four eight-GPU nodes. InfiniBand connects the nodes in both
testbeds, while the solver-efficiency experiments run on Intel Xeon Platinum 8358P CPUs. We use
Wan2.1-1.3B as the training model~\cite{wan}. The suites cover 256p--1080p images and 480p--1080p
videos, informed by the training settings of Wan and HunyuanVideo and large-scale video datasets
including OpenVid-1M and Koala-36M~\cite{wan,hunyuanvideo,openvid,koala36m}. We construct high- and
low-load suites with 10-second videos, both spanning 35--60\% video-token share at per-GPU token
budgets of 350K and 125K tokens, respectively, and repeat the same video-token-share sweep with
15-second videos. For the 15-second workloads, the per-GPU token budgets are 350K tokens on the
16-A800 testbed and 175K tokens on the 32-A6000 testbed.

\noindent\textbf{Implementation.}
Our Wan2.1 training prototype uses activation checkpointing and follows the DiffSynth-Studio block
structure and attention path~\cite{diffsynth}. We implement CP-SAT-ECF with
OR-Tools~\cite{perron2023cpsat} and execute its plans with the Coalesced Attention Engine.

\noindent\textbf{Baselines.}
We compare AdaptiveLoad~\cite{adaptiveload}, USP~\cite{usp}, and KnapFormer~\cite{knapformer}. Because
AdaptiveLoad's implementation is unavailable, we reproduce its compute-aware DP policy and
fit its load model on our testbed. We adapt xDiT's USP path to training with exact
backward~\cite{xdit,usp}, and run KnapFormer's original \texttt{SequenceBalancer}, workload model, and
Ulysses executor. We enforce a per-GPU memory constraint for KnapFormer and reject placements that
would cause OOM. All methods use identical models, batches, training loops, and
measurement protocols.

\subsection{Profile Accuracy}
\label{sec:new-eval-accuracy}

We evaluate the profile on 21 plan compositions. Eighteen mixed plans form a rotation
covering array: each contains one 480p, one 720p, and one 1080p video plus a
fixed set of 50 images. The 18 profiled parallelism configurations are assigned three per plan and
rotated across the resolutions, so every resolution--configuration combination is executed exactly
once. The remaining three are all-whole corner plans containing, respectively, only the 50 images,
three 480p videos plus those images, and three 720p videos plus those images.
Against GPU measurements across all 21 plans, the profile's mean absolute error (MAE) and mean
absolute percentage error (MAPE) are $2.5$~s and $3.4\%$ for step makespan and $0.4$~GiB and
$1.5\%$ for peak allocated memory.

\subsection{Solver Efficiency}

We compare the two-stage planner with the joint-placement reference, evaluating both with and
without ECF on the high-load suite. The reference applies the same placement model and two rounds
to the full batch, first minimizing its maximum profiled rank load and then split count within
the same $\epsilon$ slack. It jointly optimizes anchors and fillers and returns the plan without
Filler Packing.
Table~\ref{tab:new-solver-ablation} shows two effects. First, with ECF, the joint-placement
reference certifies every cell in 4.6--89.0~s, whereas it returns no result within 600~s without
ECF; ECF also reduces the two-stage planner solve time from 53--247~ms to 33--119~ms. Second,
across the six high-load workloads, \sysname{}'s two-stage
planner with ECF incurs at most $0.32\%$ modeled step-makespan overhead relative to the
joint-placement reference ($0.10\%$ mean) while
solving $68$--$1487\times$ faster (Table~\ref{tab:new-solver-ablation}). Both CP-SAT rounds of each ECF-enabled solver report
\texttt{OPTIMAL} in every cell: Anchor Placement for \sysname{} and full-batch placement for the
reference.
\label{sec:new-eval-planner}
\begin{table}[t]
  \centering
  \caption{Solver-formulation ablation over six high-load workloads (350K tokens per GPU; 35--60\%
  video-token share). Planner solve time measures the time to produce a complete placement and
  excludes attention execution: it ends when the joint-placement reference returns its placement
  or after Filler Packing for the two-stage planner. ECF removes only permutations of
  solver-equivalent sequences and preserves both optima. The overhead row is
  $\widehat T_{\mathrm{two}}/\widehat T_{\mathrm{joint}}-1$ for the modeled makespans of the
  returned Two-stage and Joint +ECF plans; mean-column parentheses give solve-time speedup over
  Joint +ECF.}
  \label{tab:new-solver-ablation}
  \scriptsize
  \setlength{\tabcolsep}{2pt}
  \renewcommand{\arraystretch}{1.0}
  \begin{tabular*}{\columnwidth}{@{\extracolsep{\fill}}lccccccl@{}}
    \toprule
    Planner solve time (s) & V35 & V40 & V45 & V50 & V55 & V60 & mean \\
    \midrule
    Joint & \multicolumn{7}{c}{No result within 600~s} \\
    Joint +ECF & 8.1 & 5.5 & 4.6 & 63.8 & 7.9 & 89.0 & 29.8 \\
    Two-stage & 0.053 & 0.075 & 0.107 & 0.171 & 0.231 & 0.247 & 0.147 ($203\times$) \\
    {\bfseries\boldmath Two-stage +ECF} & {\bfseries\boldmath 0.033} & {\bfseries\boldmath 0.039} &
    {\bfseries\boldmath 0.037} & {\bfseries\boldmath 0.043} & {\bfseries\boldmath 0.116} &
    {\bfseries\boldmath 0.119} & {\bfseries\boldmath 0.065 ($458\times$)} \\
    \midrule
    Step-makespan overhead & 0.00\% & 0.00\% & 0.02\% & 0.00\% & 0.32\% & 0.24\% & 0.10\% \\
    \bottomrule
  \end{tabular*}
\end{table}
\subsection{End-to-End Performance}
\label{sec:new-eval-homogeneous}

\begin{figure}[t]
  \centering
  \begin{subfigure}{\columnwidth}
    \centering
    \includegraphics[width=\linewidth]{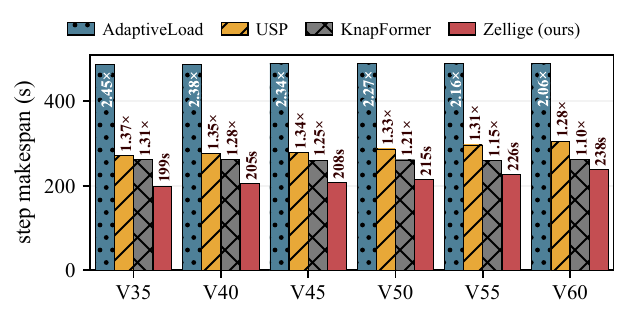}
    \caption{High-load suite (350K tokens per GPU).}
    \label{fig:end2end-share-sweep}
  \end{subfigure}
  \begin{subfigure}{\columnwidth}
    \centering
    \includegraphics[width=\linewidth,trim=3.6pt 0 3.0pt 0,clip]{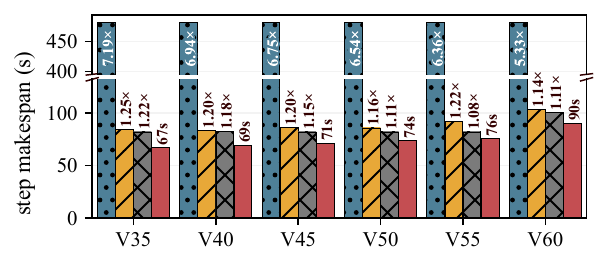}
    \caption{Low-load suite (125K tokens per GPU).}
    \label{fig:end2end-125k-control}
  \end{subfigure}
  \caption{End-to-end step makespan on eight A800 GPUs. Each panel contains six workloads spanning
  35--60\% video-token share. Annotations give \sysname{}'s speedup over each baseline; \sysname{}'s
  bars show step makespan in seconds. USP reports its best factorization and KnapFormer its best feasible disjoint-group layout
  per workload; panel~(b) uses a
  broken y-axis to keep all methods readable.}
  \label{fig:end2end-load-suites}
\end{figure}

We measure full training steps, including forward, backward, and optimizer updates, for the
10-second suites on one eight-GPU node of the A800 testbed and the 15-second workloads on both
multi-node testbeds. For
each workload, we report the best measured USP and KnapFormer results across all available
factorizations and feasible disjoint-group layouts, respectively.

\sysname{} is fastest in all twelve workloads. In Fig.~\ref{fig:end2end-load-suites}(a),
\sysname{}'s speedup over KnapFormer is
$1.10$--$1.31\times$ ($1.22\times$ mean); its mean speedups over USP and AdaptiveLoad are
$1.33\times$ and $2.28\times$. In Fig.~\ref{fig:end2end-load-suites}(b), \sysname{}'s speedup over
KnapFormer is $1.08$--$1.22\times$ ($1.14\times$ mean), while its speedup over AdaptiveLoad is
$5.33$--$7.19\times$. At the same video-token shares, the low-load suite's lower token budget makes
each indivisible video a larger fraction of the ideal per-rank workload and therefore a more severe
straggler. USP instead
pays the communication cost of splitting every sequence, while KnapFormer remains constrained by its
preconfigured disjoint groups. \sysname{} avoids these bottlenecks by splitting only compute-dominant sequences
and co-locating whole fillers on their ranks.

\begin{figure}[t]
  \centering
  \includegraphics[width=\columnwidth]{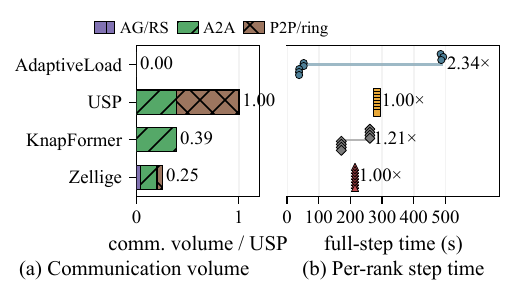}
  \caption{Measured resource breakdown for the high-load V50 workload on eight A800 GPUs
  (10-second videos; 350K tokens per GPU) over one full checkpointed training step.
  Panel~(a) shows attention data transferred between GPUs by collective type,
  normalized to USP; panel~(b) shows the eight per-rank full-step times for each method, annotated with the
  max-to-average ratio.}
  \label{fig:end2end-resource-breakdown}
\end{figure}

For the high-load V50 case in Fig.~\ref{fig:end2end-resource-breakdown}, \sysname{}
transfers $25.4\%$ as much attention data as USP and $64.9\%$ as much as
KnapFormer (panel~(a)), while matching USP's $1.00\times$
max-to-average ratio of per-rank full-step times and improving on KnapFormer's $1.21\times$
(panel~(b)). AdaptiveLoad transfers no cross-rank attention data but
has the largest per-rank time imbalance, with a $2.34\times$ max-to-average ratio because long
videos remain indivisible.

\begin{figure}[t]
  \centering
  \begin{subfigure}{\columnwidth}
    \centering
    \includegraphics[width=\linewidth]{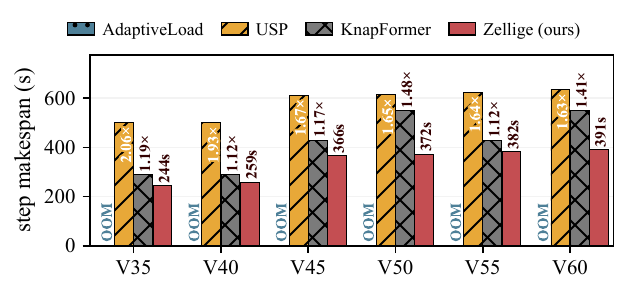}
    \caption{16 A800 GPUs.}
    \label{fig:new-eval-scalability-n16}
  \end{subfigure}
  \begin{subfigure}{\columnwidth}
    \centering
    \includegraphics[width=\linewidth]{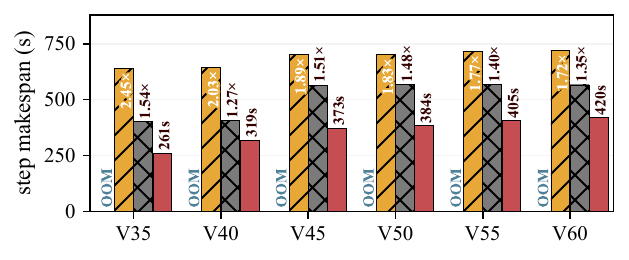}
    \caption{32 A6000 GPUs.}
    \label{fig:new-eval-scalability-a6000-n32}
  \end{subfigure}
  \caption{End-to-end step makespan for six workloads with 15-second videos spanning
  35--60\% video-token share at device-specific
  high-load budgets. Annotations give \sysname{}'s speedup
  over each baseline; KnapFormer uses its best feasible disjoint-group layout per workload. AdaptiveLoad OOMs in all
  six workloads of both panels.}
  \label{fig:new-eval-scalability}
\end{figure}

For workloads with 15-second videos on 16 A800 GPUs, \sysname{}'s speedup over KnapFormer is
$1.12$--$1.48\times$ ($1.25\times$ mean), and its speedup over USP is
$1.63$--$2.06\times$ ($1.76\times$ mean) (Fig.~\ref{fig:new-eval-scalability}(a)).
On 32 A6000 GPUs, \sysname{}'s speedup over KnapFormer is
$1.27$--$1.54\times$ ($1.42\times$ mean), and its speedup over USP is
$1.72$--$2.45\times$ ($1.95\times$ mean) (Fig.~\ref{fig:new-eval-scalability}(b)).
AdaptiveLoad OOMs on both deployments because the memory footprint of an indivisible 15-second
video exceeds the per-GPU memory cap (Fig.~\ref{fig:new-eval-scalability}).

\section{Related Work}
\label{sec:related-new}

AdaptiveLoad uses compute-aware DP~\cite{adaptiveload}, USP applies Ulysses/ring-based CP to every
sequence~\cite{dsp,ulysses,ringattention,usp}, and KnapFormer combines them through
preconfigured disjoint rank groups~\cite{knapformer}. It keeps sequences whole or splits them, but
each inherits its group's parallelism configuration and rank set. Other diffusion systems optimize inference or
model-stage placement, not per-sequence attention placement~\cite{xdit,pipedit,pulse}.

Long-context LLM systems also explore sequence partitioning and placement across
GPUs. FlexSP solves a bandwidth-aware per-minibatch MILP over disjoint Ulysses groups~\cite{flexsp}. HotSPa groups
sequences by length and sequentially switches among group-specific
parallel strategies rather than concurrently executing per-sequence placement options on overlapping rank
sets~\cite{hotspa}. ByteScale supports dynamic per-sequence meshes and packing, but without offload it
splits sequences only when required by memory and uses ring attention~\cite{bytescale}. DCP~\cite{dcp} permits
fine-grained block co-location and minimizes the hypergraph connectivity-minus-one
communication-volume objective~\cite{kahypar} under balance constraints, targeting total traffic rather than step
makespan.

\section{Conclusion}
\label{sec:conclusion-new}

We proved that disjoint-group placement faces a fundamental tradeoff between inter-group load
imbalance and intra-group communication redundancy.
To address this tradeoff in mixed image-video DiT training, we presented \sysname{}, a moldable
sequence placement system that selects each sequence's parallelism configuration and participating
ranks.
\sysname{} consists of a Hardware Profiler, a two-stage planner that balances compute-heavy anchors
and packs lighter fillers into the remaining capacity, and a Coalesced Attention Engine that
efficiently executes whole sequences alongside distributed-attention shards.
The two-stage planner solves each batch in 33--119~ms, significantly faster than a joint-placement
reference that optimizes all sequences together, while their modeled makespans differ by at most
$0.32\%$.
In end-to-end evaluations, \sysname{} outperforms KnapFormer by $1.12$--$1.48\times$ on 16 A800 GPUs
and $1.27$--$1.54\times$ on 32 A6000 GPUs.

\bibliographystyle{IEEEtran_etal}
\bibliography{references}

\end{document}